\documentstyle[11pt,epsfig]{book}

\def\mHalf{m_{1/2}}
\def\mZero{m_0}
\def\AZero{A_0}
\def\NO{{\tilde\chi^0_1}} 
\def\NT{{\tilde\chi^0_2}} 
\def\lN{{l_{\rm n}}}
\def\lF{{l_{\rm f}}}
\def\qN{{q_{\rm n}}}
\def\qF{{q_{\rm f}}}
\def\gl{{\tilde g}}

\def\sle{{\tilde l}}
\def\lR{{\tilde l_R} }
\def\qL{{\tilde q_L}}

\def\mNO{{m_\NO}}
\def\mlR{{m_\lR}}
\def\mNT{{m_\NT}}
\def\mqL{{m_\qL}}
\def\mgl{{m_\gl}}
\def\GeV{{\rm GeV}}

\def\mll{m_{ll}}
\def\mqlN{m_{q\lN}}
\def\mqlF{m_{q\lF}}
\def\mqll{m_{qll}}

\def\sNO{{m_\NO^2}}
\def\slR{{m_\lR^2}}
\def\sNT{{m_\NT^2}}
\def\sqL{{m_\qL^2}}

\def\mqlLow{m_{ql(\low)}}
\def\mqlHigh{m_{ql(\high)}}

\def\itB{\it}

\def\low{{\rm low}}
\def\high{{\rm high}}
\def\equal{{\rm eq}}

\def\max{{\rm max}}
\def\maxmqlHigh{m^\max_{ql(\high)}}
\def\palpha{{($\alpha$)}}
\def\pbeta{{($\beta$)}}

\def\maxmqlLow{m^\max_{ql(\low)}}
\def\maxmqlHigh{m^\max_{ql(\high)}}

\def\av#1{\langle #1 \rangle}
\def\spcA{\mbox{\hspace{1.2ex}}}

\begin{document}
\chapter*{ }
\vskip -5truecm
\vskip 15truemm

\centerline {\bf DETERMINING MASSES OF SUPERSYMMETRIC PARTICLES}  
\vskip 10truemm
\centerline {\bf B.~K.~Gjelsten$^1$, D.~J.~Miller$^2$, P.~Osland$^3$ }  
\vskip 5truemm
\centerline{${}^1$ {
Laboratory for High Energy Physics, University of Bern, CH-3012 Bern,}}
\centerline{Switzerland}
\centerline {${}^2$ {Department of Physics and Astronomy,
University of Glasgow,}}
\centerline {Glasgow G12 8QQ, U.K.}
\centerline {${}^3$ {Department of Physics,
University of Bergen, N-5007 Bergen, Norway}}

{\renewcommand{\thefootnote}{}
\footnote {\hskip -5mm Presented at the {\it Southeastern European
Workshop Challenges Beyond the Standard Model}, 19-23 May 2005, Vrnjacka
Banja, Serbia}
}
\setcounter{page}{0}   
\thispagestyle{empty}

\markboth{B.~K.~Gjelsten, D.~J.~Miller, P.~Osland}
{DETERMINING MASSES OF SUPERSYMMETRIC PARTICLES}  

\begin{quote}
\small{\bf Abstract.} {\it
If supersymmetric particles are produced at the Large Hadron Collider
it becomes very important not only to identify them,
but also to determine their masses
with the highest possible precision, since this may lead to an understanding
of the SUSY-breaking mechanism and the physics at some higher scale.
We here report on studies of how such mass measurements are obtained,
and how the precision can be optimized.}
\end{quote}

\begin{quote}
\small{\bf Key words:} {\it SUSY, BSM, MSSM}
\end{quote}
\vskip 5truemm

\centerline {\sc 1. INTRODUCTION}

Supersymmetry~\cite{Fayet:1976cr} has attracted a lot of attention as a
natural extension of the Standard Model of particle physics.  Not only does it
solve the hierarchy problem, it also has many other attractive features, the
most celebrated of which are that it provides a {\it natural} mechanism for
generating the Higgs potential which breaks the electroweak
symmetry~\cite{Inoue:1982pi} and that it supplies a good candidate for cold
dark matter~\cite{Goldberg:1983nd}.  Furthermore, if it is to be relevant in
solving the hierarchy problem it must exhibit experimental consequences at the
TeV-scale, and therefore can be tested by experiments at the Large Hadron
Collider (LHC).

If supersymmetric particles are produced at the LHC, it will become important
to identify them and accurately measure their masses. Only an accurate
determination of the supersymmetric particle masses and couplings will allow
us to determine the low energy soft supersymmetry breaking parameters.  It is
hoped that extrapolation of these masses and couplings to high energies will
provide an insight into the mechanism of supersymmetry breaking and, more
generally, physics at the GUT scale~\cite{Allanach:2004ud}. 

Here we will discuss supersymmetric mass measurements with reference to one
particular model of supersymmetry breaking, minimal super-gravity
(mSUGRA)~\cite{Inoue:1982pi,Chamseddine:jx}. In this model, the supersymmetry
is broken by the interaction of new particles at high energy which are only
linked to the usual particles by gravitational interactions; this new sector
of physics is often referred to as the {\it hidden sector}. 
At the GUT scale, the scalar
supersymmetric particles are assumed to have a common mass, $\mZero$, while
the gauginos have a common mass $\mHalf$. The trilinear couplings are also
taken to be universal at the GUT scale and denoted $\AZero$.

Furthermore, in this model R-parity is conserved. 
Thus, supersymmetric particles are produced in pairs, and the lightest one
(from each decay chain) escapes detection.
While this provides a very distinctive {\it missing energy} signature,
it complicates the measuring of masses at the LHC since decays 
cannot be fully reconstructed.

Instead, mass measurements rely on continuous mass distributions
of decay products which attain extrema for certain configurations of
the particle momenta that are unambiguously determined by the masses
of initial, intermediate and final particles involved.
These relations may often be inverted
to give the masses of unstable particles. 

\begin{figure}[htb]
\refstepcounter{figure}
\label{Fig:1}
\addtocounter{figure}{-1}
\begin{center}
\setlength{\unitlength}{1cm}
\begin{picture}(8,2.5)
\put(-1.,0){
\mbox{\epsfysize=3.5cm\epsffile{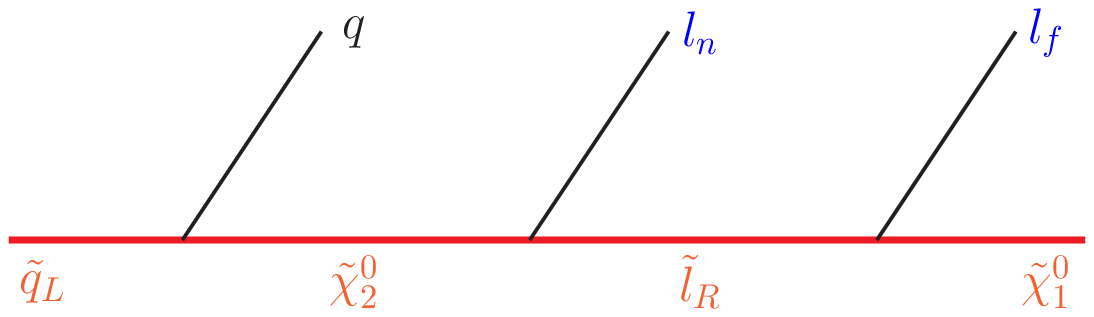}}}
\end{picture}
\vspace*{-4mm}
\caption{Typical cascade decay chain.}
\end{center}
\end{figure}

We here report on how supersymmetric mass measurements can be made
\cite{Gjelsten:2004ki,Gjelsten:2005aw}
by examining the mass distribution endpoints or `edges' of the long decay
chains (see Fig.~\ref{Fig:1})
\begin{equation}
\label{Eq:sq-chain}
\tilde q \to \NT q \to \sle \lN q \to \NO \lF \lN q
\end{equation}
and
\begin{equation}
\label{Eq:gl-chain}
\tilde g \to \tilde q\qN \to \NT \qF\qN \to \sle \lN \qF\qN \to \NO \lF \lN
\qF\qN,
\end{equation}
where $\tilde g$ denotes the gluino, $\tilde q$ a squark, $\sle$ a slepton,
whereas $\NT$ and $\NO$ are the two lightest neutralinos.
For easier reference a subscript is given to the quarks and leptons 
when needed (``n'' for ``near'' and ``f'' for ``far''). 

For the decay chain (\ref{Eq:sq-chain}) the following four invariants can be
defined:
\begin{equation}
\mll, \qquad\mqlN, \qquad\mqlF, \qquad\mqll,
\label{Eq:masses-1}
\end{equation}
where the subscripts are left out when there is no ambiguity. 
For the longer chain (\ref{Eq:gl-chain}) seven more invariants can be 
defined, but for simplicity we will here keep mostly to the first case, 
as adding the gluino in any case involves no additional complications on the
conceptual level. 
\bigskip

\centerline {\sc 2. INVARIANT MASS DISTRIBUTIONS}

In practice, one will usually not be able to distinguish the ``near'' and the
``far'' leptons of Fig.~\ref{Fig:1} and Eq.~(\ref{Eq:masses-1}). Therefore,
two alternative distributions are defined:
\begin{equation}
\mqlLow=\min(\mqlN,\mqlF), \quad
\mqlHigh=\max(\mqlN,\mqlF).
\end{equation}

For the Snowmass mSUGRA benchmark point SPS~1a~\palpha\ 
\cite{Allanach:2002nj,Gjelsten:2004ki}, 
the electroweak-scale masses are given as
\begin{equation}
\label{Eq:sps1a-alpha}
\{\mqL,\ \mNT,\ \mlR,\ \mNO\}
=\{537.2,\ 176.8,\ 143.0,\ 96.1\}~\GeV.
\end{equation}
For these mass values, the invariant distributions for
\begin{equation}
\mll, \qquad\mqlLow, \qquad\mqlHigh, \qquad\mqll
\label{Eq:masses-2}
\end{equation}
are given in Fig.~\ref{Fig:2}.
For comparison, we also show the corresponding distributions for
the point SPS~1a \pbeta\ \cite{Gjelsten:2004ki}, a related mass scenario, 
also on the SPS~1a line \cite{Allanach:2002nj}, for which the masses are
\begin{equation}
\label{Eq:sps1a-beta}
\{\mqL,\ \mNT,\ \mlR,\ \mNO\}
=\{826.3,\ 299.1,\ 221.9,\ 161.0\}~\GeV,
\end{equation}

We see that while the dilepton mass distribution has a simple triangular
shape, and $\mqlLow$ for SPS~1a has a smooth rounded shape, 
the other distributions have several abrupt changes in slope.
The reason for this is that they are composite, with different functions
describing the different intervals. In fact, these functions have recently
been worked out analytically \cite{Raklev}.
The great variety of these functional forms represents a complication 
to the experimental determination of the over-all endpoints.  
\begin{figure}[htb]
\refstepcounter{figure}
\label{Fig:2}
\addtocounter{figure}{-1}
\begin{center}
\setlength{\unitlength}{1cm}
\begin{picture}(8,6.5)
\put(-3.2,3.2){
\mbox{\epsfysize=4.5cm\epsffile{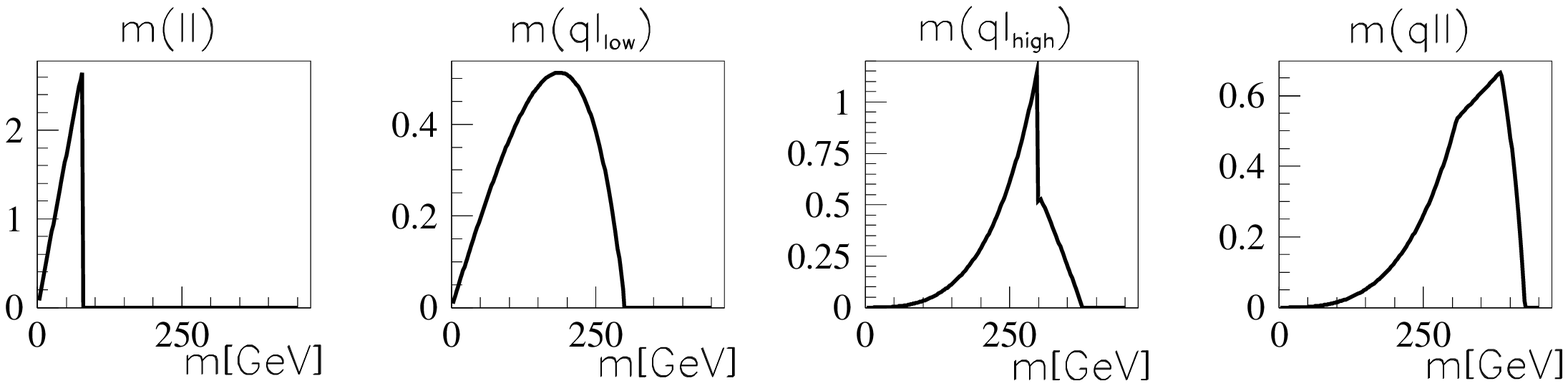}}}
\put(-3.2,0){
\mbox{\epsfysize=4.5cm\epsffile{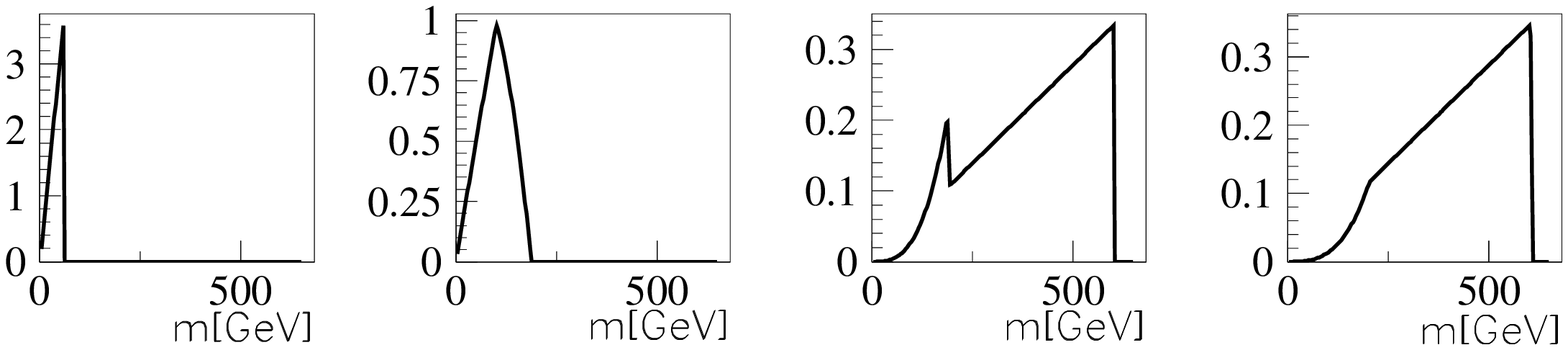}}}
\end{picture}
\vspace*{-6mm}
\caption{Invariant mass distributions for 
the SPS~1a \palpha\ \cite{Allanach:2002nj,Gjelsten:2004ki} (upper row)
and SPS~1a \pbeta\ \cite{Gjelsten:2004ki} (lower row) benchmark points.}
\end{center}
\end{figure}

Most of the endpoints are given by composite functions with different 
expressions for different mass constellations. 
For example, the endpoints of $\maxmqlLow$ and $\maxmqlHigh$ are given by
\cite{Allanach:2000kt,Gjelsten:2004ki}:
\begin{equation}
\label{Eq:mql_low-mql_high}
\big(\maxmqlLow,\maxmqlHigh\big) 
= \left\{
\begin{array}{llc}
\big(m^\max_{q\lN},m^\max_{q\lF}\big) &
\quad {\rm for } \quad &{\rm Case~1} \\[2mm]
\big(m^\max_{ql(\equal)},m^\max_{q\lF}\big) &
\quad {\rm for } \quad &{\rm Case~2} \\[2mm]
\big(m^\max_{ql(\equal)},m^\max_{q\lN}\big) &
\quad {\rm for } \quad &{\rm Case~3}
\end{array} \right\} 
\label{eq:m_ql}
\end{equation}
with
\def\sqL{{m_\qL^2}}
\begin{eqnarray}
\big(m^\max_{q \lN}\big)^2
&=& \big(\sqL-\sNT\big)\big(\sNT-\slR\big)/\sNT
\label{Eq:edge-qlN} \nonumber \\[2mm]
\left(m^\max_{q \lF}\right)^2
&=& \big(\sqL-\sNT\big)\big(\slR-\sNO\big)/\slR
\label{Eq:edge-qlF} \nonumber \\[2mm]
\big(m^\max_{ql(\equal)}\big)^2 &=&
\big(\sqL-\sNT\big)\big(\slR-\sNO\big)/\big(2\slR-\sNO\big)
\label{Eq:edge-ql-equal} 
\end{eqnarray}
and the different cases given by the linear and quadratic ratios of
the underlying (unknown) masses:
\begin{eqnarray}
{\rm Case~1}:\quad &2\slR > \sNO+\sNT > 2\mNO\mNT \nonumber \\
{\rm Case~2}:\quad &\sNO+\sNT > 2\slR > 2\mNO\mNT \nonumber \\
{\rm Case~3}:\quad &\sNO+\sNT > 2\mNO\mNT > 2\slR 
\end{eqnarray}

These methods for determining the masses of supersymmetric particles have been
extensively studied in the past \cite{Baer:1995va}.  Our studies
\cite{Gjelsten:2004ki,Gjelsten:2005aw,Gjelsten:2005sv} extend the earlier
works by (i) discussing theoretical distributions which arise for different
mass scenarios (examples are shown in Fig.~\ref{Fig:2}), (ii) providing
inversion formulas 
(by using the endpoints of the four distributions (\ref{Eq:masses-2}) the edge
formulas can be solved for the masses $\mqL$, $\mlR$, $\mNT$ and $\mNO$),
(iii) discussing ambiguities and complications related to the composite nature
of the endpoint expressions, 
(iv) extension of the method to include a gluino at the head of the chain. 
\bigskip

\centerline {\sc 3. Results}

In order to estimate the precision that may be obtained at the LHC, for the
actual masses $\mgl$, $\mqL$, $\mlR$, $\mNT$ and $\mNO$, we have investigated 
the decay chains (\ref{Eq:sq-chain}) and (\ref{Eq:gl-chain}) for two mass
scenarios, SPS~1a~\palpha\ of Eq.~(\ref{Eq:sps1a-alpha}) 
and SPS~1a~\pbeta\ of Eq.~(\ref{Eq:sps1a-beta}). 

The latter has higher masses. This would be an advantage for the kinematics,
but higher masses also lead to lower cross sections, and thus a signal more
easily masked by the background. The simulation was performed using the Monte
Carlo program PYTHIA~6.2 \cite{PYTHIA}, and the events were then passed
through ATLFAST \cite{ATLFAST}, a simulation of the ATLAS detector.

A critical issue in this kind of analysis is the background. 
The Standard Model background can to a large extent be removed, it turns out, 
by exploiting the fact that the two leptons must be of the same flavour. 
In most cases, the SM background events which mimic the signal, 
such as $t\bar t \to b\bar bW^+W^- \to b\bar bl^+l^-\nu\bar\nu$ 
produce as many different-flavour lepton pairs ($e^\pm\mu^\mp$) 
as same-flavour lepton pairs ($e^+e^-, \mu^+\mu^-$)\footnote{
Taus are not used because of their bad experimental resolution.}. 
Since these two background selections are statistically identical, 
the same-flavour part, which is collected together with the signal, 
can be removed statistically by subtracting the different-flavour part. 
Other SUSY decays will however contribute background events that are harder to
control. The considered squark (or gluino) will typically be produced in
association with another squark or gluino, which gives rise to a similar decay
chain. This leads most importantly to an additional hard jet.  (We make here
the usual assumption that the mass difference between the squarks and the
sparticles into which they decay is large, producing on average the hardest
jets in the event.)  As it is not possible to know which of the two leading
jets belongs to the signal, one will in principle select the incorrect jet in
half of the cases or more.  This adds considerably to the background and is
the main reason why the experimental distributions tend to differ noticeably
from the original distributions.

\begin{figure}[htb]
\refstepcounter{figure}
\label{Fig:3}
\addtocounter{figure}{-1}
\begin{center}
\setlength{\unitlength}{1cm}
\begin{picture}(12.5,8.0)
\put(0.0,4.2){\mbox{\epsfysize=4.0cm\epsffile{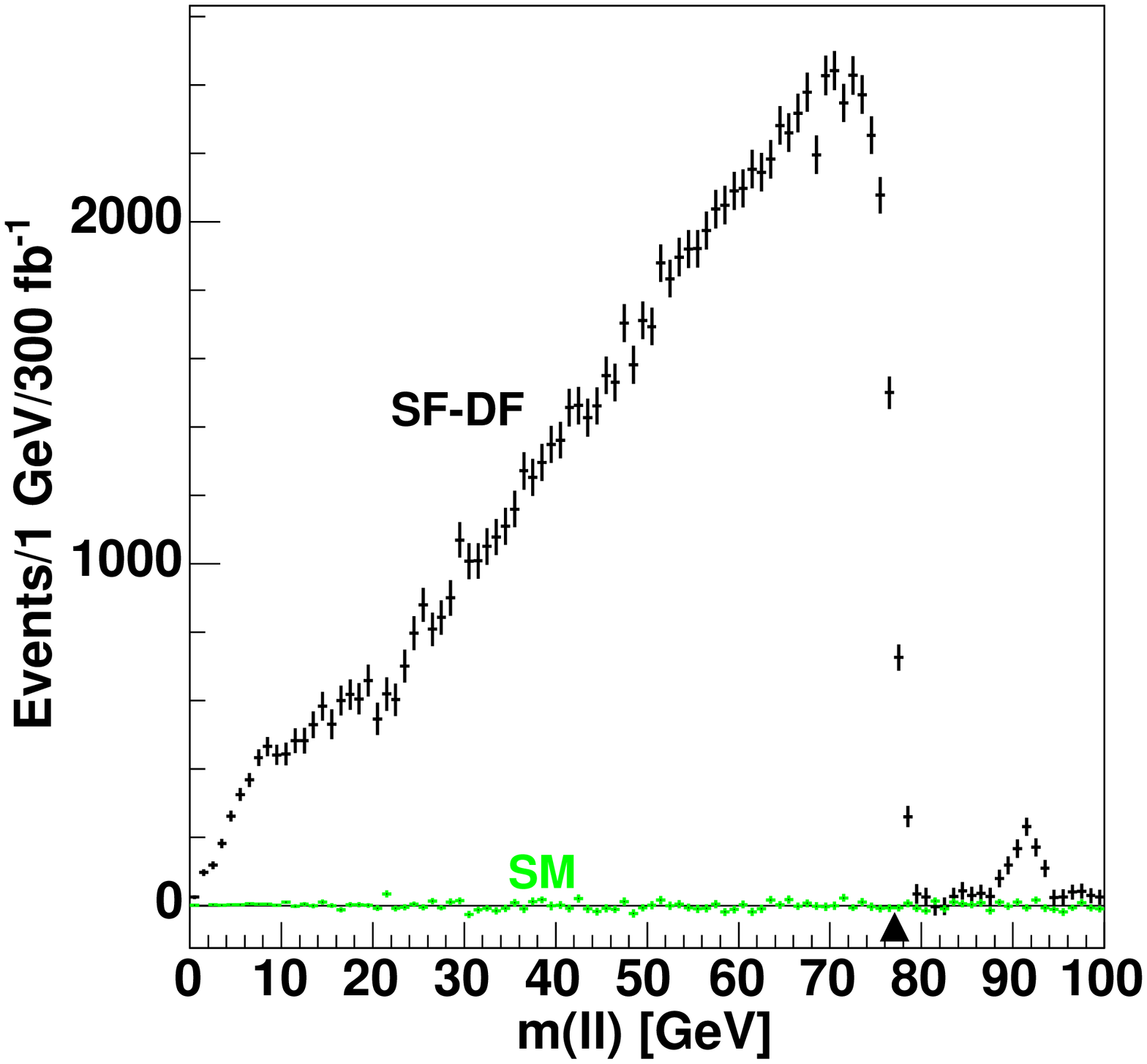}}}
\put(4.2,4.2){\mbox{\epsfysize=4.0cm\epsffile{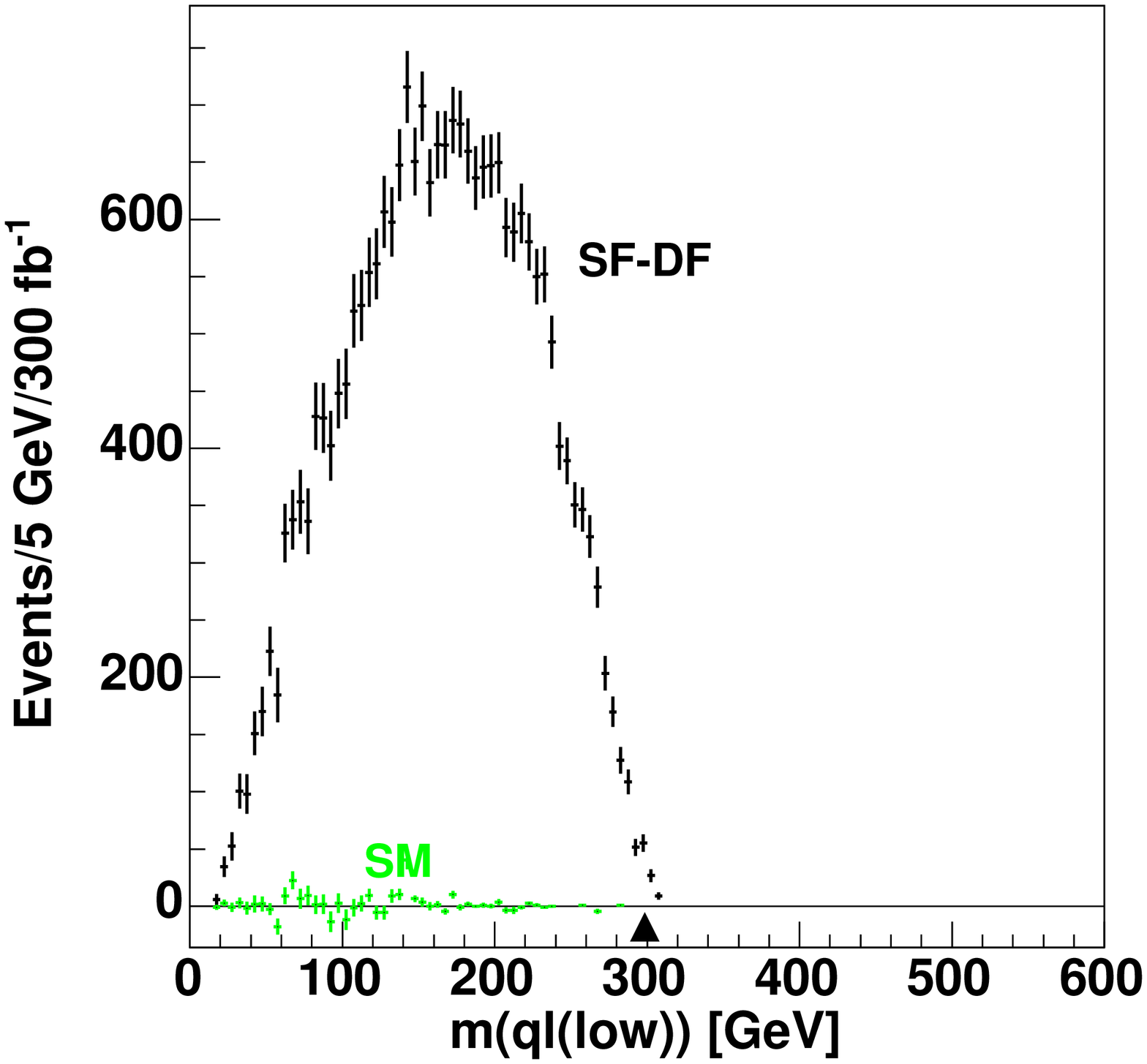}}}
\put(8.4,4.2){\mbox{\epsfysize=4.0cm\epsffile{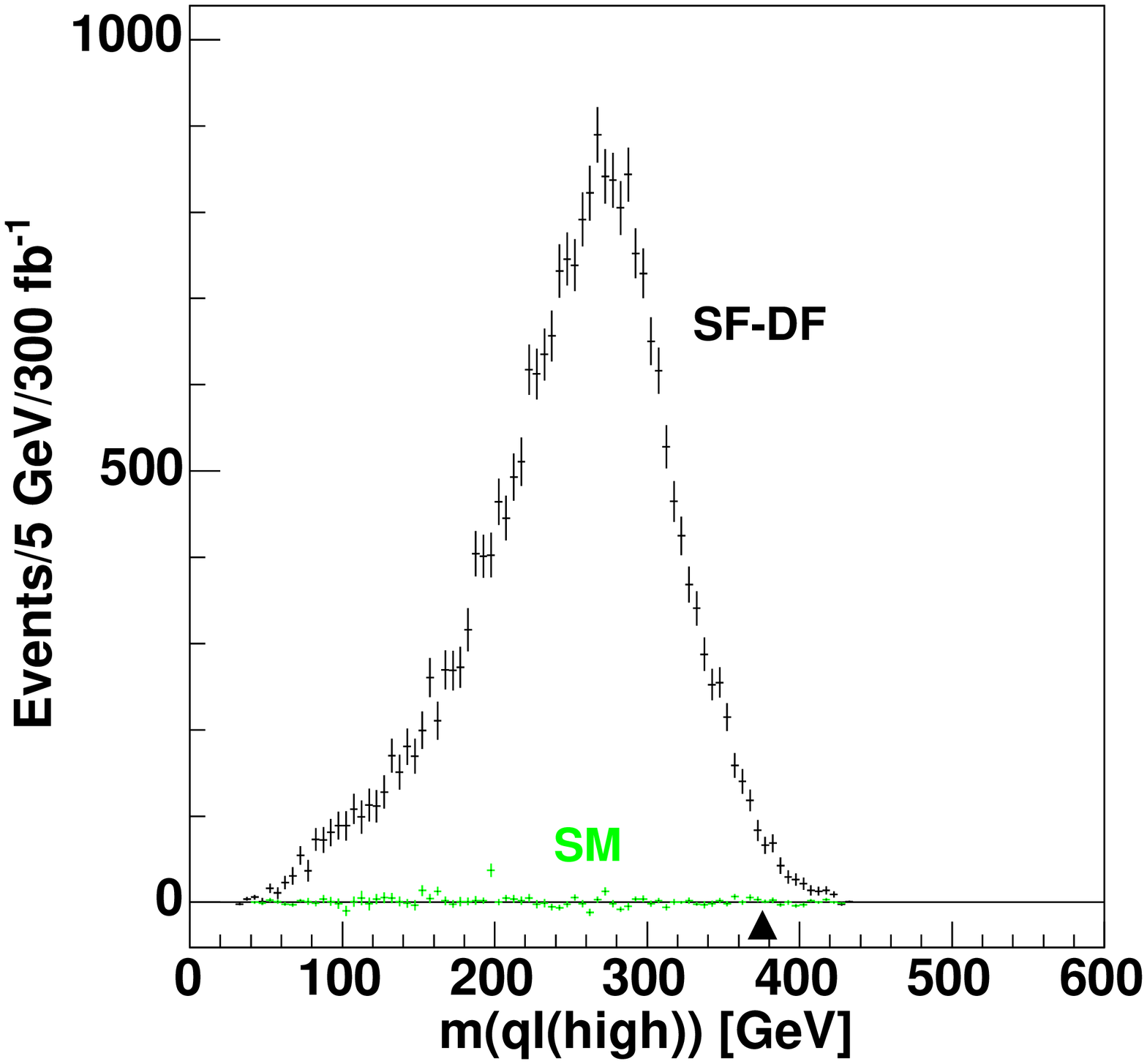}}}
\put(0.0,0.0){\mbox{\epsfysize=4.0cm\epsffile{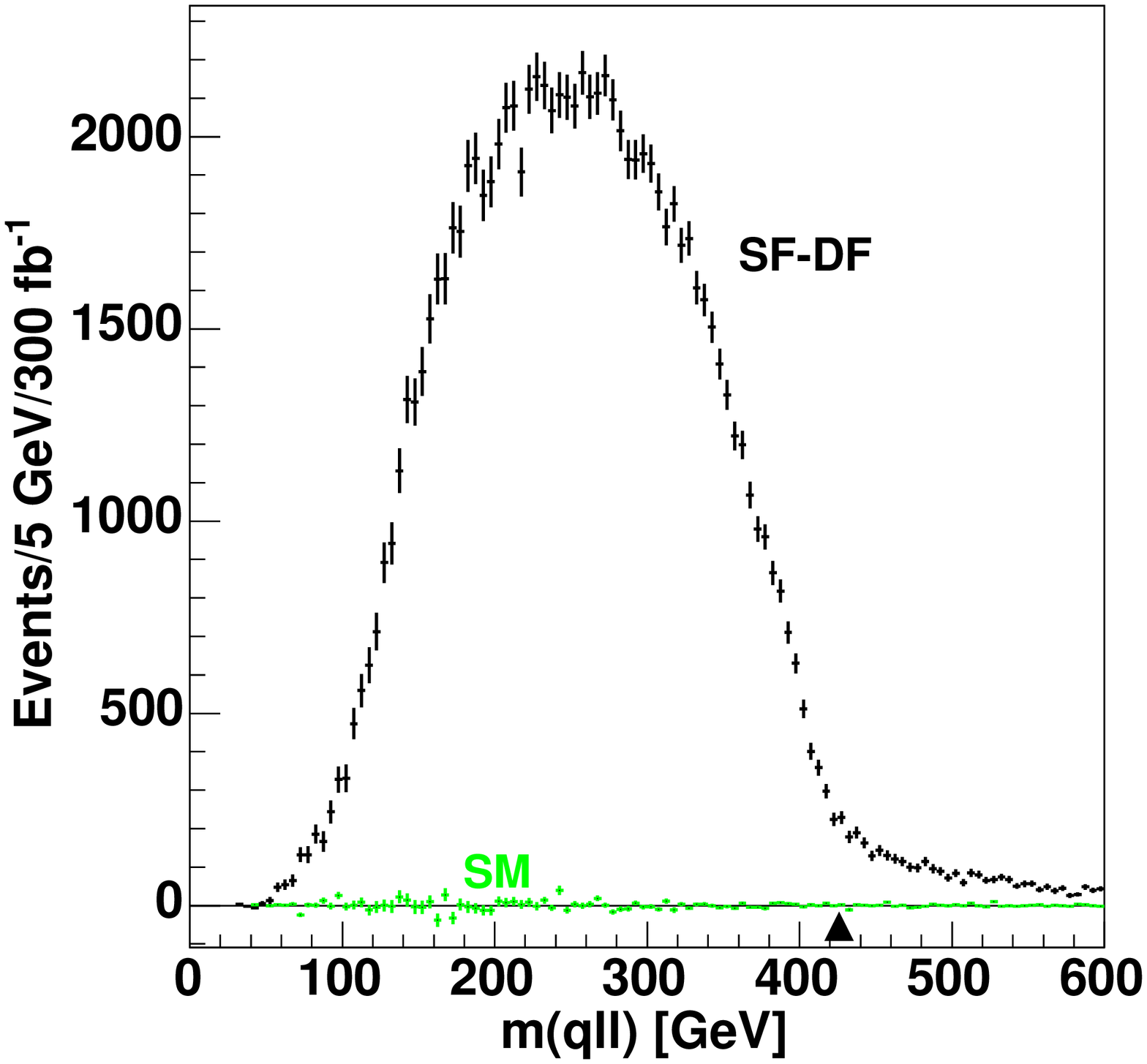}}}
\put(4.2,0.0){\mbox{\epsfysize=4.0cm\epsffile{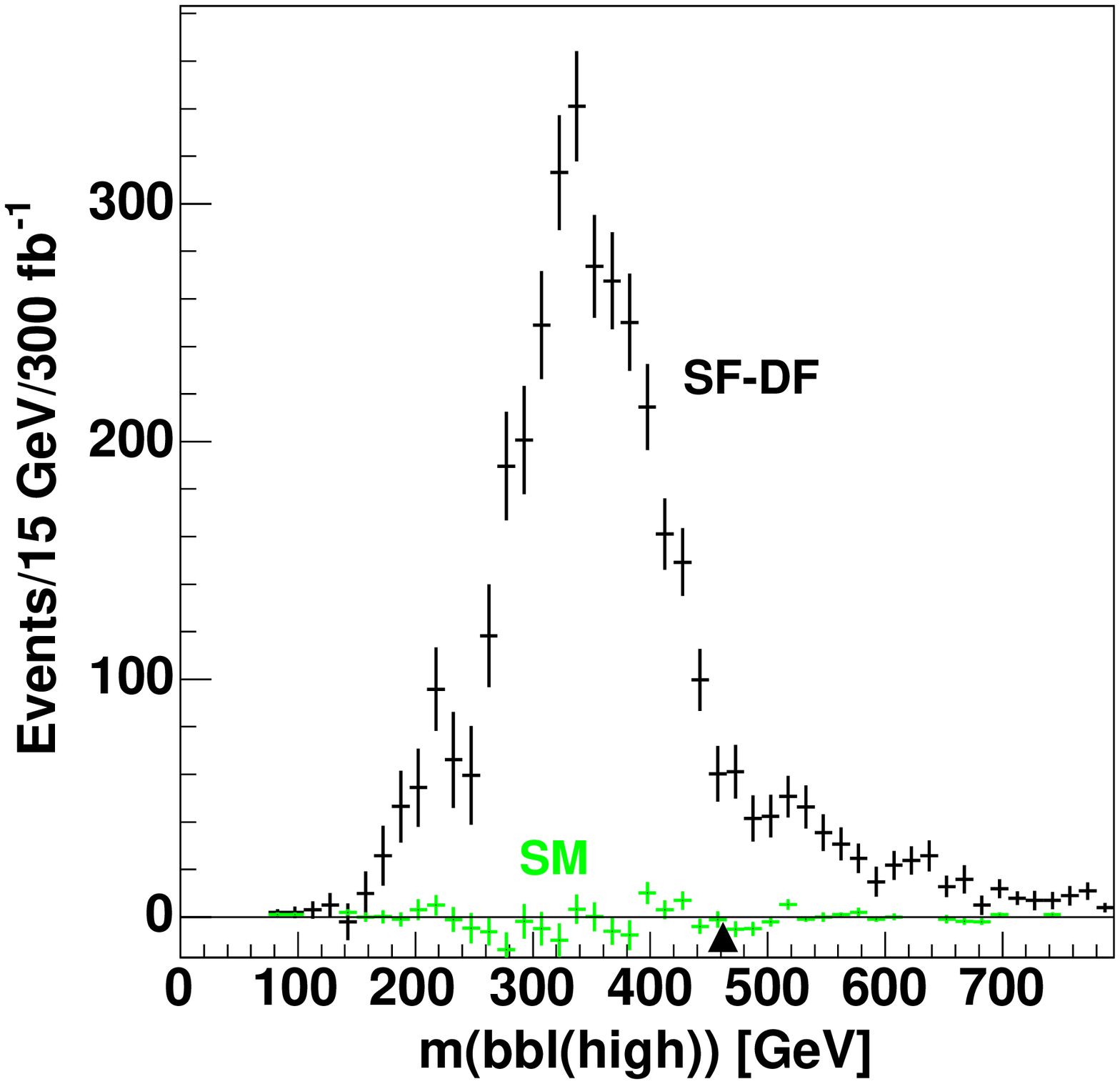}}}
\put(8.4,0.0){\mbox{\epsfysize=4.0cm\epsffile{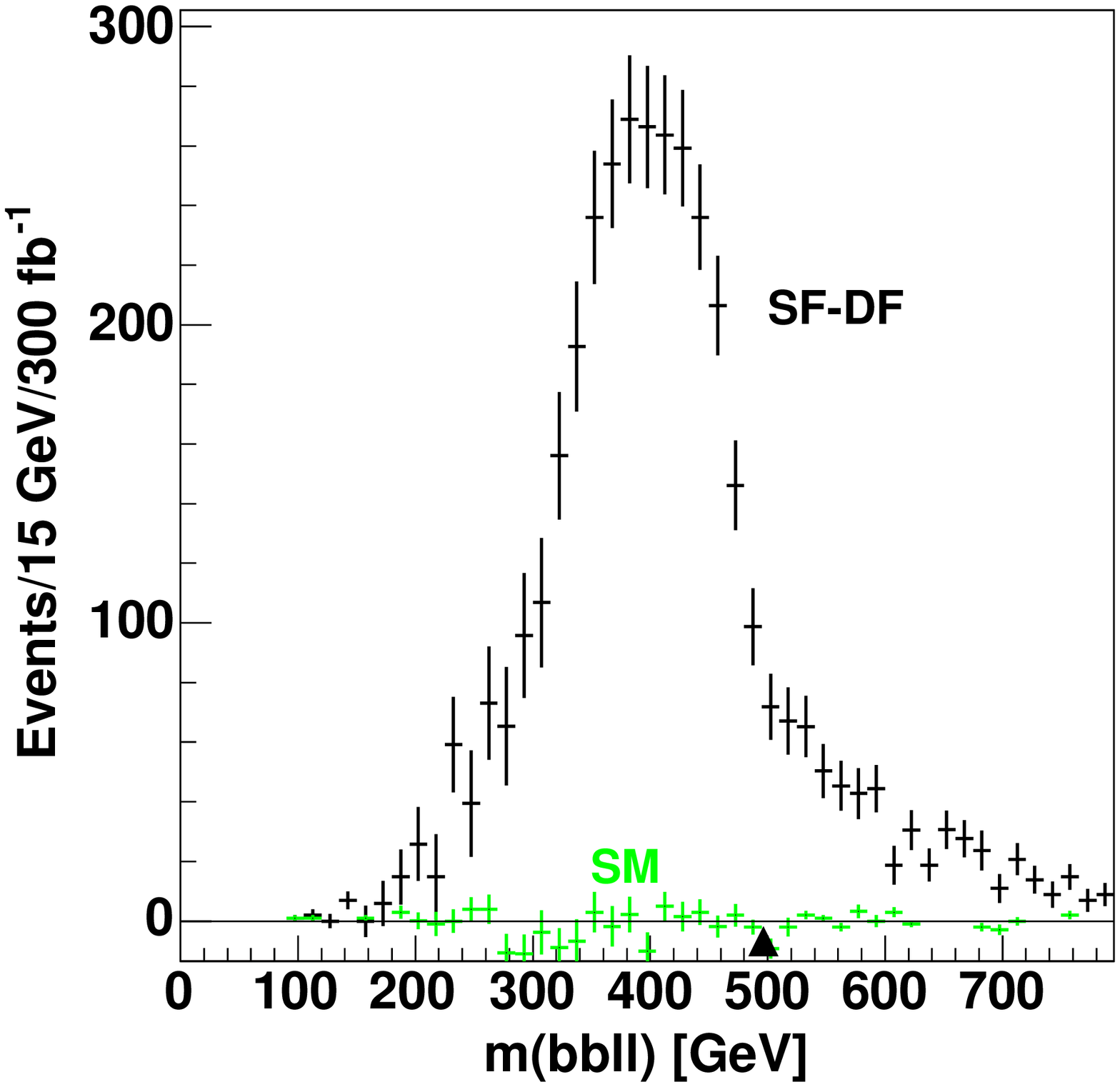}}}
\end{picture}
\vspace*{-2mm}
\caption{Some of the available experimental mass distributions for
  SPS~1a~$(\alpha)$. The triangles mark the positions of the theoretical
  endpoints.}  
\end{center}
\end{figure}

\begin{figure}[htb]
\refstepcounter{figure}
\label{Fig:4}
\addtocounter{figure}{-1}
\begin{center}
\setlength{\unitlength}{1cm}
\begin{picture}(12.5,3.8)
\put(0.0,0){\mbox{\epsfysize=4.0cm\epsffile{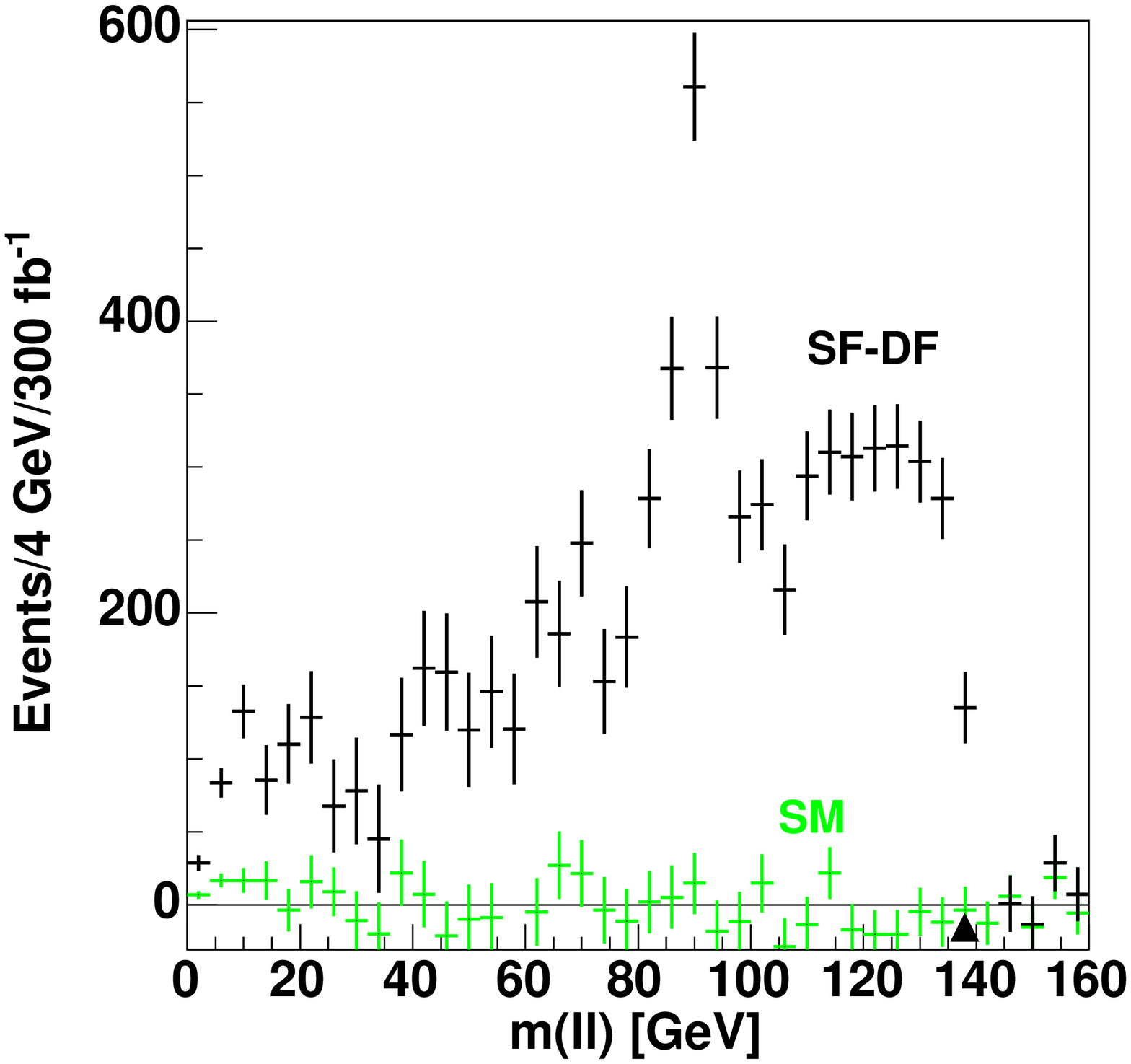}}}
\put(4.2,0){\mbox{\epsfysize=4.0cm\epsffile{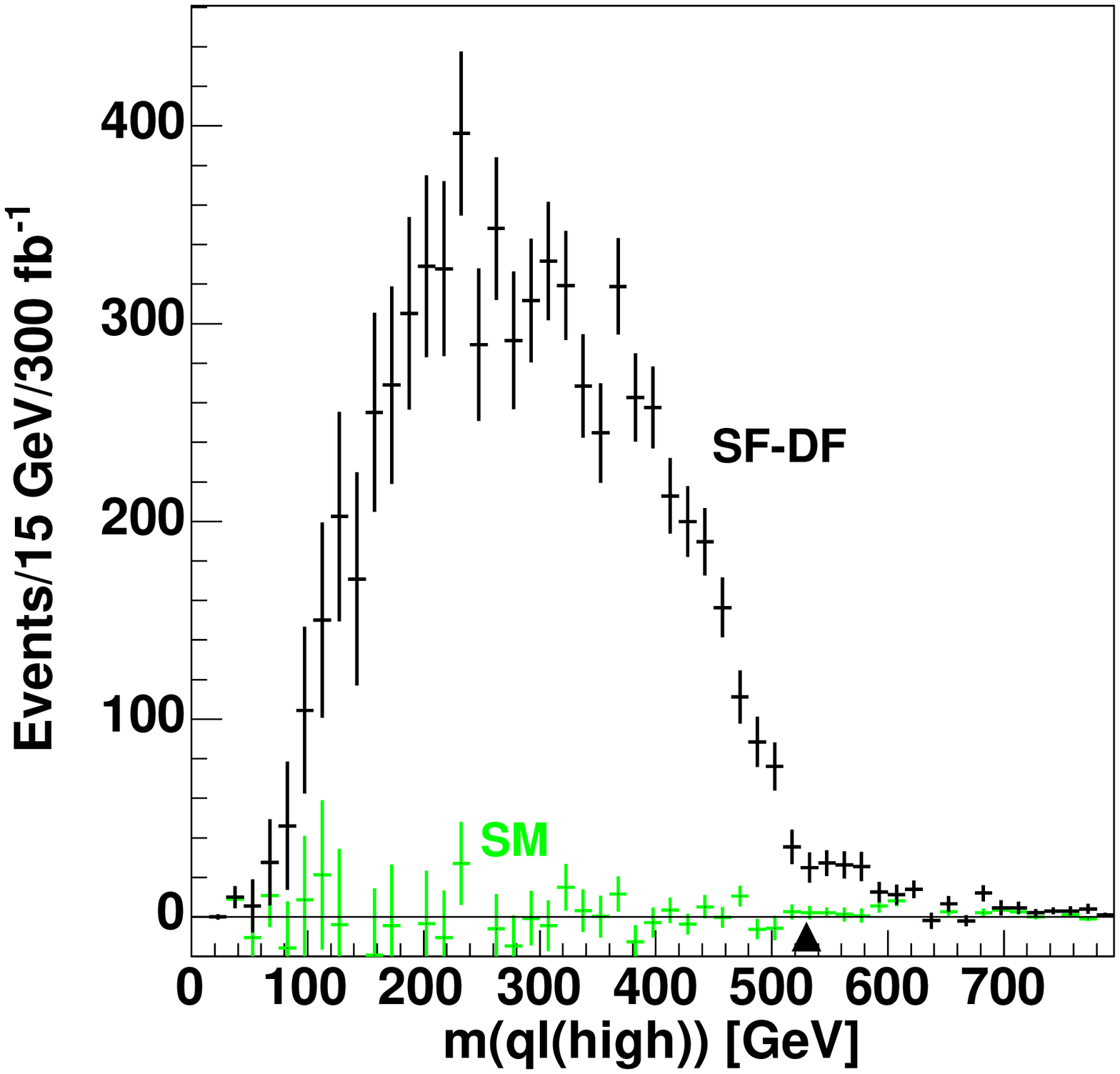}}}
\put(8.4,0){\mbox{\epsfysize=4.0cm\epsffile{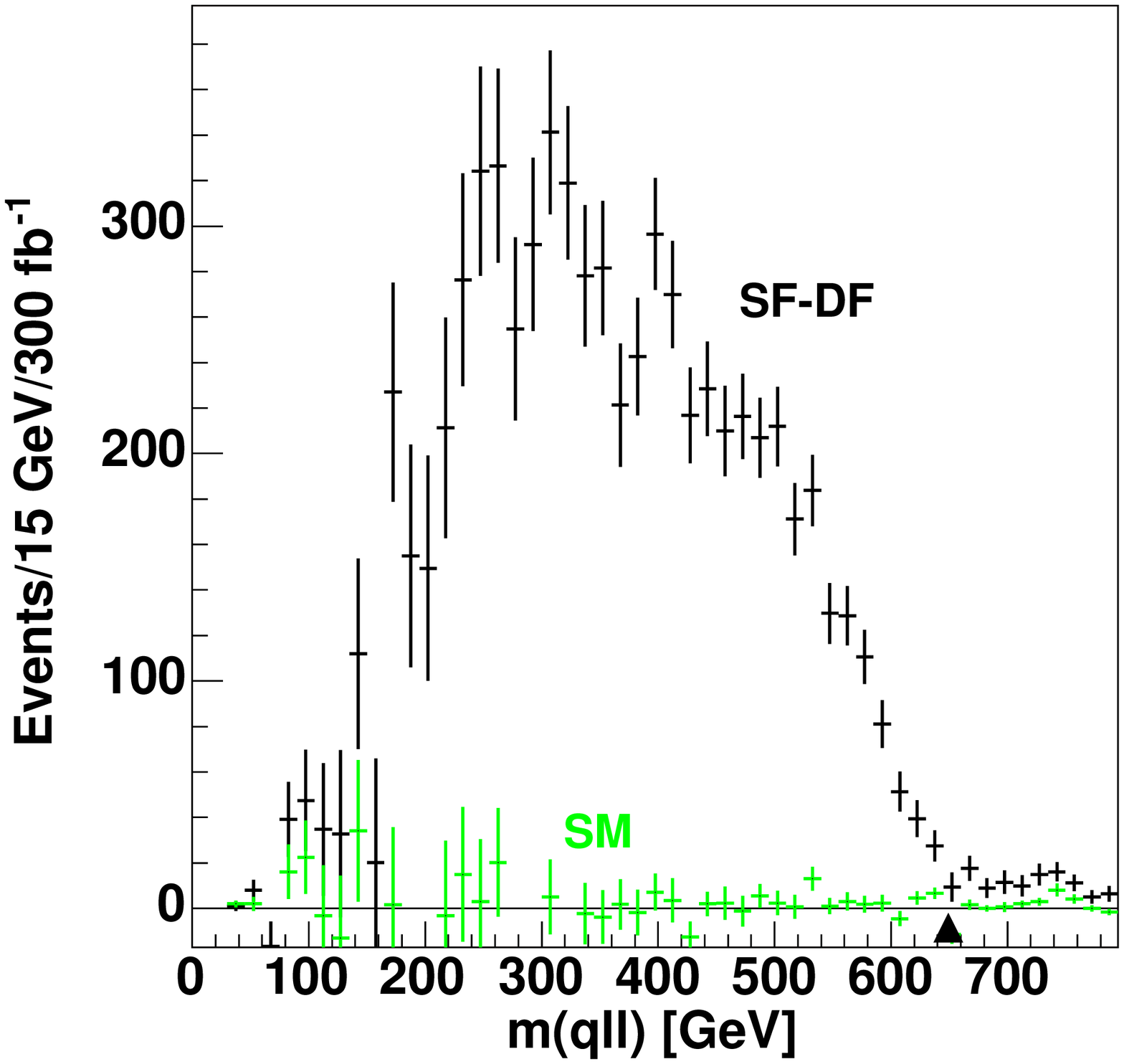}}}
\end{picture}
\vspace*{-2mm}
\caption{Some of the available experimental mass distributions for
  SPS~1a~$(\beta)$. The triangles mark the positions of the theoretical
  endpoints.}  
\end{center}
\end{figure}

For the two SUSY scenarios reported on here, a selection of the available 
different-flavour-subtracted mass distributions (`SF-DF') are shown 
in Figs.~\ref{Fig:3} and \ref{Fig:4}.  
The plots are given for  300 fb$^{-1}$, 
which is attained after three years at design luminosity. 
The Standard Model part of the samples is plotted separately in green 
(`SM'). 
Although the SM background is likely to be somewhat underestimated 
in these analyses, it is seen to have limited impact on the 
total distributions. 
The first four distributions of Fig.~\ref{Fig:3} correspond to the 
upper set of distributions in Fig.~\ref{Fig:2}. While the experimental 
version of the $m_{qll}$ distribution no longer resembles its theory
counterpart, appropriate cuts amend this situation for the two $m_{ql}$
distributions, although at the price of reduced 
statistics. The two last distributions are constructed from events with 
two $b$-tagged jets 
and access the gluino part of the decay chain, allowing for the 
determination of the gluino mass. 

Although there are noticeable smearing and background effects in 
all the distributions, the edges are quite pronounced and point 
in the close vicinity of the theoretical endpoint. 
This is also true for the SPS~1a~($\beta$) distributions, where the 
signal cross-section is some 25 times lower. 
Measuring the distribution endpoints is in itself a troublesome task. Apart
from the statistical uncertainty, there is a (yet) largely uncontrolled
systematic error from the signal and background hypothesis made in the actual
fitting.  Attempts at improvement are being pursued, and will be much more so
if SUSY is discovered at the LHC.  In our investigation we have therefore
taken the optimistic short-cut and assumed that this systematic error will be
dominated by the controllable statistical and jet energy scale errors.

Furthermore, 
rather than obtaining the SUSY masses for the specific endpoint measurements 
found from the distributions in Figs.~\ref{Fig:3} and \ref{Fig:4} as well as 
similar distributions not shown here, these endpoints, or more precisely 
their statistical errors, were used as indicative of what the LHC will 
be able to do, and 
an {\em ensemble} of typical LHC experiments was generated. 
For each experiment the SUSY masses were found from a numerical $\chi^2$ fit.  
See \cite{Gjelsten:2004ki,Allanach:2000kt} for details.

While the endpoint measurements provide us with sufficient information to 
obtain the masses, inherent to the method are two complicating effects. 
First, the endpoints relate to {\it mass differences} much more than to the 
masses themselves, hence the errors on mass differences are much smaller 
than on the masses. 
For the lighter masses the precision for the two quantities 
can typically differ by around an order of magnitude. 
This must be considered a limitation of the method, since what enters 
further application are usually the masses, not mass differences. 
The most familiar example is the input to the RGE equations which 
on the basis of the SUSY masses allow us to make statements on the 
GUT-scale physics  \cite{Allanach:2004ud}. 
Another example is cross-section considerations. 
Second, there is a high probability that several sets of masses give 
the same set of endpoints. 
If the number of masses is equal to the number of measured endpoints, 
this is exactly true. If more endpoint measurements are available, 
the system becomes overconstrained in favour of the correct solution, 
but due to measurement uncertainties, more than one solution 
must still often be considered. The relevant measure for comparing minima, 
is their difference in $\chi^2$ value for the mass fit. 
If this difference is small, both minima are at the same level of
compatibility with the endpoint measurements and must be accepted. 
In order to choose one rather than the other, more information 
is required than what the endpoints alone provide. 
(For a discussion of these ambiguities, see \cite{Gjelsten:2005sv}.)


In the case of SPS~1a~($\alpha$) there are two competing solutions, 
the nominal one, which is in mass region {\it(1,1)}\footnote{See
\cite{Gjelsten:2004ki} for the definition of the mass regions.}, and a second
one in region {\it(1,2)}.  
If we accept solutions for which the difference in
$\chi^2$ down to the global minimum is less than 1 (3),
there is a 12\% (30\%) probability that the system has two solutions.  The
probability that the nominal-region solution appears is very high, but there
is a substantial admixture of the {\it(1,2)} solution.  
In Table~\ref{table:1}
the resulting masses for the two solution types are shown together with the
nominal values (``Nom''). The masses are given first, then mass differences,
with $\mNO$ selected as reference point.  The better precision of the mass
differences is apparent, especially for the lighter masses.

\begin{table}[htb]{ 
\caption{SPS~1a~($\alpha$).
Nominal masses, ``Nom'',
ensemble means, $\av{m}$, and root-mean-square distances
from the mean, $\sigma$, all in GeV.\label{table:1}} }
%
\begin{center}
\begin{tabular}{|c|r|rr|rr|}
\hline
&  &\multicolumn{2}{|c|}{\itB(1,1)} &\multicolumn{2}{|c|} {\itB(1,2)}  \\
& Nom & $\av{m}$ & $\sigma\hspace{1.5ex}$ & $\av{m}$ 
& $\sigma \hspace{1.5ex}$  \\
\hline
$\mNO    $ &   96.1\spcA  &   96.3\spcA &  3.8\spcA &   85.3 &  3.4\spcA \\
$\mlR    $ &  143.0\spcA  &  143.2\spcA &  3.8\spcA &  130.4 &  3.7\spcA \\
$\mNT    $ &  176.8\spcA  &  177.0\spcA &  3.7\spcA &  165.5 &  3.4\spcA \\
$\mqL    $ &  537.2\spcA  &  537.5\spcA &  6.0\spcA &  523.5 &  5.0\spcA \\
$\mgl    $ &  595.2\spcA  &  595.5\spcA &  7.2\spcA &  582.5 &  6.8\spcA \\
\hline
$\mlR-\mNO$ &   46.92  &   46.93 &  0.28 &   45.11 &  0.72 \\ 
$\mNT-\mNO$ &   80.77  &   80.77 &  0.18 &   80.19 &  0.29 \\ 
$\mqL-\mNO$ &  441.2\spcA  &  441.2\spcA &  3.1\spcA &  438.0\spcA &  2.8\spcA  \\
$\mgl-\mNO$ &  499.1\spcA  &  499.2\spcA &  5.6\spcA &  497.0\spcA &  5.4\spcA  \\
\hline
\end{tabular}\end{center}
\end{table}

In the case of SPS~1a~($\beta$) three minima are competing, the nominal 
solution, a second solution with similar masses and a third solution 
with much higher masses. Again, the nominal solution is 
available in most cases, but the admixture from the other one or two 
is significant. The preeminence of mass differences over masses, in terms 
of precision, is once more confirmed. 
Numbers can be found in \cite{Gjelsten:2004ki}.

Both the intrinsic ``weaknesses'' of the endpoint method discussed above 
involve an uncertainty of the SUSY mass scale. 
A Linear Collider measurement of the LSP mass effectively sets this scale, 
so combining the measurements from the LHC with those from a Linear Collider 
improves drastically the accuracy on the SUSY masses \cite{lhc-lc}. 
\bigskip

P. O. would like to thank the organizers of the {\it Southeastern European
Workshop Challenges Beyond the Standard Model}, 19-23 May 2005, Vrnjacka
Banja, Serbia for the invitation to a very interesting meeting.
This work has been performed partly within the ATLAS Collaboration,
and we thank collaboration members for helpful discussions.
We have made use of the physics analysis framework and tools
which are the result of collaboration-wide efforts.
This research has been supported in part by the Research Council of Norway.

\renewcommand {\bibname} {\normalsize \sc References}

{}

\end{document}